\documentclass[superscriptaddress,showpacs,twocolumn]{revtex4}
\usepackage{amssymb}
\usepackage[tbtags]{amsmath}
\usepackage{graphicx}
\usepackage{epsfig,graphicx,times}

\begin{document}

\date{\today}
\title{Quantum computation with Josephson-qubits by using a
current-biased information bus}
\author{L.F. Wei}
\affiliation{Frontier Research System, The Institute of Physical and Chemical Research
(RIKEN), Wako-shi, Saitama, 351-0198, Japan}
\affiliation{Institute of Quantum Optics and Quantum Information, Department of Physics,
Shanghai Jiaotong University, Shanghai 200030, P.R. China }
\author{ Yu-xi Liu}
\affiliation{Frontier Research System, The Institute of Physical and Chemical Research
(RIKEN), Wako-shi, Saitama, 351-0198, Japan}
\author{Franco Nori}
\affiliation{Frontier Research System, The Institute of Physical and Chemical Research
(RIKEN), Wako-shi, Saitama, 351-0198, Japan}
\affiliation{Center for Theoretical Physics, Physics Department, Center for the Study of
Complex Systems, The University of Michigan, Ann Arbor, Michigan 48109-1120}

\begin{abstract}
We propose an effective scheme for manipulating quantum
information stored in a superconducting nanocircuit. The Josephson
qubits are coupled via their separate interactions with an
information bus, a large current-biased Josephson junction treated
as an oscillator with adjustable frequency. The bus is
sequentially coupled to only one qubit at a time. Distant
Josephson qubits without any direct interaction can be indirectly
coupled with each other by independently interacting with the bus
sequentially, via exciting/de-exciting vibrational quanta in the
bus. This is a superconducting analog of the successful ion trap
experiments on quantum computing. Our approach differs from
previous schemes that simultaneously coupled two qubits to the
bus, as opposed to their sequential coupling considered here. The
significant quantum logic gates can be realized by using these
tunable and selective couplings. The decoherence properties of the
proposed quantum system are analyzed within the Bloch-Redfield
formalism. Numerical estimations of certain important experimental
parameters are provided.

\vspace{0.3cm}
\end{abstract}

 \maketitle

\vspace{-0.2cm}

\section{Introduction.}

The coherent manipulation of quantum states for realizing certain
potential applications, e.g, quantum computation and quantum
communication, is attracting considerable interest \cite{NC00}. In
principle, any two-state quantum system works as a qubit, the
fundamental unit of quantum information. However, only a few real
physical systems have worked as qubits, because of requirements of
a long coherent time and operability. Among various physical
realizations, such as ions traps (see, e.g.,
\cite{CZ95,Wei02,Wineland03}), QED cavities (see, e.g.,
\cite{SW95,Raimond01}), quantum dots (see, e.g.,~\cite{Hu03,SX01})
and NMR~(see, e.g.,~\cite{Chuang97,Lieven01}) etc.,
superconductors with Josephson junctions offer one of the most
promising platforms for realizing quantum computation (see,
e.g.,~\cite{Makhlin01,Tsai99,Mooij99,Vion02,Tsai03,Averin03,Makhlin99,
You02,Falci03,You03,Blais03,Ramos01,Wallraff03,Martinis02,Yu02,Martinis03,
Berkley03, Strauch03,Fan01,Liu04}). The nonlinearity of Josephson
junctions can be used to produce controllable qubits. Also,
circuits with Josephson junctions combine the intrinsic coherence
of the macroscopic quantum state and the possibility to control
its quantum dynamics by using voltage and magnetic flux pulses. In
addition, present-day technologies of integration allow scaling to
large and complex circuits. Recent experiments have demonstrated
quantum coherent dynamics in the time domain in both single-qubit
(see, e.g.,~\cite{Tsai99,Mooij99,Vion02}) and two-qubit Josephson
systems ~\cite{Tsai03}.

There are two basic types of Josephson systems used to implement
qubits: charge qubits~\cite{Tsai99} and flux
qubits~\cite{Mooij99}, depending on the ratio of two
characteristic energies: the charging energy $E_{C}$ and the
Josephson energy $E_{J} $. The charge qubit is a Cooper-pair box
with a small Josephson coupling energy, $E_J\ll E_C$, and a well
defined number of Cooper pairs is well defined. The flux qubit
operates in another extreme limit, where $E_{J}\gg E_{C}$ and the
phase is well defined. A ``quantronium" circuit operating in the
intermediate regime of the former two has also been
proposed~\cite{Vion02}. Voltage-biased superconducting quantum
interference devices (SQUIDs), which work in the charge regime and
with controllable Josephson energies, form the SQUID-based charge
qubits that we will consider in this work. Our results can be
extended to flux and flux-charge qubits.

The key ingredient for computational speedup in quantum
computation is entanglement, a property that does not exist in
classical physics. Thus, manipulating coupled qubits plays a
central role in quantum information processing (QIP).
Heisenberg-type qubit-couplings are common for the usual solid
state QIP systems, e.g., the real spin states of the electrons in
quantum dots~\cite{Hu03,SX01}. However, the interbit couplings for
Josephson junctions involve Ising-type interactions, as
superconducting qubits with two macroscopic quantum states provide
pseudo-spin-1/2 states. Recently, either the current-current
interaction, by connecting to a common inductor, or the
charge-charge coupling, via sharing a common capacitor, have been
proposed to directly couple two Josephson charge qubits: the $i$th
and $j$th ones. These interactions implement
$\sigma^{(i)}_z\otimes\sigma^{(j)}_z$--type~\cite{Tsai03,Averin03},
$\sigma^{(i)}_y\otimes\sigma^{(j)}_y$--type~\cite{Makhlin99}, and
the $\sigma^{(i)}_x\otimes\sigma^{(j)}_x$--type~\cite{You02} Ising
couplings, respectively. Compared to the single-qubit operations,
the two-qubit operations based on these second-order interactions
are more sensitive to the environment. Thus, quantum decoherence
can be more problematic. In addition, capacitive coupling between
qubits is not easily tunable~\cite{Tsai03}. Thus adjusting the
physical parameters for realizing two-qubit operation is not easy.
In order to ensure that the quanta of the relevant $LC$ oscillator
is not excited during the desired quantum operations, the time
scales of manipulation in the inductively coupled circuit should
be much slower than the eigenfrequency of the
$LC$-circuit~\cite{Makhlin99}.

Alternatively, the Josephson qubits may also be coupled together
by sequentially interacting with a data bus, instead of
simultaneously. This is similar to the techniques used for trapped
ions~\cite{CZ95,Wei02}, wherein the trapped ions are entangled by
exciting and de-exciting quanta of their shared center-of-mass
vibrational mode (i.e., the data bus). This scheme allows for
faster two-qubit operations and
possesses longer decoherent times. In fact, an externally connected $LC$%
-resonator \cite{Falci03} and a cavity QED mode~\cite{You03} were
chosen as alternative data buses. However, it is not always easy
to control all the physical properties, such as the
eigenfrequencies and decoherence, of these data buses.

A large (e.g., up to $10\mu$m) current-biased Josephson junction
(CBJJ)~\cite{Blais03} is very suitable to act as information bus
for coupling Josephson qubits. This because: i) the CBJJ is an
easily fabricated device~\cite{Ramos01} and may provide more
effective immunities to both charge and flux noise; ii) due to its
large junction capacitance, the CBJJ can enable to be capacitively
coupled over relatively long distances; iii) the quantum
properties, e.g., quantum transitions between the junction energy
levels, of the current-biased Josephson junction are well
established \cite{Clarke88,Wallraff03}; and iv) its eigenfrequency
can be controlled by adjusting the applied bias-current. In fact,
a CBJJ itself can be an experimentally realizable qubit, as
demonstrated by the recent observations of Rabi oscillations in
them~\cite{Martinis02,Yu02}. Two logic states of such a qubit are
encoded by the two lowest zero-voltage metastable quantum energy
levels of the CBJJ. The decoherent properties of this CBJJ-qubit
were discussed in detail in~\cite{Martinis03}. Experimentally, the
entangled macroscopic quantum states in two CBJJ-qubits coupled by
a capacitor were created~\cite{Berkley03}. Also, by numerical
integration of the time-dependent Schr\"odinger equation, a full
dynamical simulation of two-qubit quantum logic gates between two
capacitively coupled CBJJ-qubits was given in~\cite{Strauch03}.

In this paper, we propose a convenient scheme to selectively
couple two Josephson charge-qubits. Here, a large CBJJ acts only
as the information bus for transferring the quantum information
between the qubits. Thus, hereafter the CBJJ will not be a qubit,
as
in~\cite{Blais03,Martinis02,Yu02,Martinis03,Berkley03,Strauch03}.
Two chosen distant SQUID-based charge qubits can be indirectly
coupled by sequentially interacting these with the bus. This
coupling method provides a repeatable way to generate entangled
states, and thus can implement elementary quantum logic gates
between arbitrarily selected qubits. Our proposal shares some
features with the circuits proposed in
\cite{Makhlin99,Blais03,Falci03,You02}, but also has significant
differences. Our proposal might be more amenable to experimental
verification.

The outline of the paper is as follows. In Sec. II we propose a
superconducting nanocircuit with a CBJJ acting as the data bus,
and investigate its elemental quantum dynamics. The bus is biased
by a $dc$ current and is assumed to interact with only one qubit
at a time. There is no direct interaction between qubits.
Therefore, the elemental operations in this circuit consist of: i)
the free evolution of the single qubit, ii) the free evolution of
the bus, and iii) the coherent dynamics for a single qubit coupled
to the bus. In Sec. III we show how to realize the elemental logic
gates in the proposed nanocircuit: the single-qubit rotations by
properly switching on/off the applied gate voltage and external
flux, and the two-qubit operations by letting them couple
sequentially to the bus. The vibrational quanta of the bus is
excited/absorded during the qubit-bus interactions. In Sec. IV we
analyze the decoherent properties of the present qubit-bus
interaction within the Bloch-Redfield formalism~\cite{AK64}, and
give some numerical estimates for experimental implementations.
Conclusions and some discussions are given in Sec. V.

\section{A superconducting nanocircuit and its elementary quantum evolutions.%
}

The circuit considered here is sketched in Fig. 1. It consists of
$N$ voltage-biased  SQUIDs connected to a large CBJJ. The $k$th
($k=1,2,...,N$) qubit consists of a gate electrode of capacitance
$C_{g_k}$ and a single-Cooper-pair box with two ultrasmall
Josephson junctions of capacitance $C^0_{J_k}$ and Josephson
energy $E^0_{J_k}$, forming a DC-SQUID ring. The inductances of
these DC-SQUID rings are assumed to be
very small and can be neglected. The SQUIDs work in the charge regime with $%
k_BT\ll E_J\ll E_C\ll\Delta$, in order to suppress quasi-particle
tunneling or excitation. Here, $k_B$, $\Delta,\,E_C$, $T$, and
$E_J$ are the Boltzmann constant, the superconducting gap,
charging energy, temperature, and the Josephson coupling energy,
respectively.

The connected large CBJJ biased  by a dc current works in the {\it
phase regime} with $E_J\gg E_C$. It acts as a tunable anharmonic
$LC$-resonator with a nonuniform level spacing and works as a data
bus for transferring quantum information between the chosen
qubits.
The mechanism for manipulating quantum information in the present
approach is different from that
in~\cite{Makhlin99,Blais03,Falci03,You02}, although the circuit
proposed here might seem similar to those there. The differences
are:

(1) a large CBJJ, instead of
$LC$-oscillator~\cite{Makhlin99,Falci03,You02} formed by the
externally connected inductance $L$ and the capacitances in
circuit, works as the data bus;

(2) we modulate the applied external flux, instead of the
bias-current~\cite{Blais03}, to realize the perfect
coupling/decoupling between the chosen qubit and the bus; and
especially

(3) the free evolution of the bus during the operational delays
will be utilized for the first time to control the dynamical
phases for implementing the expected quantum gates.
\begin{figure}[tbp]
\vspace{2cm} \includegraphics[width=12.6cm]{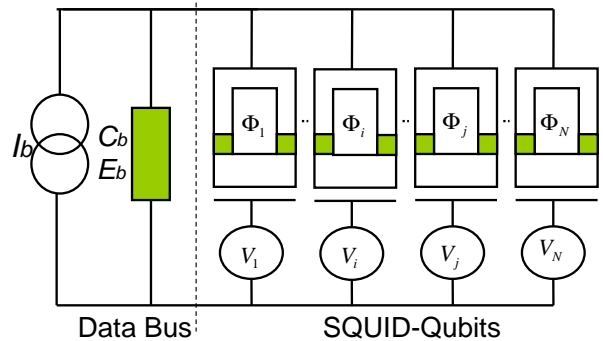}
\vspace{-6cm} \caption{SQUID-based charge qubits coupled via a
large CBJJ.} \label{fig_qubit_coupler}
\end{figure}

The Hamiltonian for the present circuit can be written as
\begin{eqnarray}
\hat{H}=\sum_{k=1}^N\left[\frac{2e^2}{C_k}\left(\hat{n}_k-n_{g_k}%
\right)^2- E_{J_k}\cos\left(\hat{\theta}_{k}-\frac{C_{g_k}}{C_k}\hat{\theta}%
_{b}\right)\right]+\hat{H}_r,
\end{eqnarray}
with
\begin{equation}
\hat{H}_r=\frac{(2\pi\hat{p}_{\,b}/\Phi_0)^2}{2\,\tilde{C}_b}-E_b\cos\hat{\theta}_b
-\frac{\Phi_0\,I_b}{2\pi}\hat{\theta}_b.
\end{equation}
Here, $n_{g_k}=C_{g_k}V_k/(2e),\,C_k=C_{g_k}+C_{J_k},\,C_{J_k}=2C^0_{J_k}$, $%
\tilde{C} _b=C_b+\sum_{k=1}^NC_{J_k}C_{g_k}/C_k$,\,$E_{J_k}=2E^0_{J_k}\cos(%
\pi\Phi_k/\Phi_0)$, and $\theta_k=(\theta_{k_2}+\theta_{k_1})/2$ with $%
\theta_{k_1}$ and $\theta_{k_2}$ being the phase drops across two small
Josephson junctions in the $k$th qubit, respectively. Also, $C_{g_k}$, $%
\Phi_0$, $\Phi_k$, and $V_k$ are the gate capacitance, flux
quantum, external flux, and gate voltage applied to the $k$th
qubit, respectively. Correspondingly, $C_b,\,\theta_b$,\,$E_b$,
and $I_b$ are the capacitance, phase drops, Josephson energy, and
the bias-current of the large CBJJ, respectively. Above, the
number operator $\hat{n}_k $ of excess Cooper-pair
charges in the superconducting island and the phase operator $\hat{\theta}%
_{k}$ of the order parameter of the $k$th charge qubit are a pair
of canonical variables and satisfy the commutation relation:
\begin{equation*}
[\,\hat{\theta}_k, \hat{n}_k\,]=i.
\end{equation*}
The operators $\theta_b$ and $\hat{p}_{\,b}$ are another pair of
canonical variables and satisfy the commutation relation:
\begin{equation*}
[\,\hat{\theta}_b, \hat{p}_{\,b}\,]=i\hbar,
\end{equation*}
with $2\pi\,p_{\,b}/\Phi_0=2\,n_b\,e$ representing the charge difference
across the CBJJ.

The CBJJ works in the phase regime. Thus, $E_{C_b}=e^2/(2\tilde{C}%
_b)\ll E_{b}$ and the quantum motion ruled by the Hamiltonian
$\hat{H}_r$ equals to that of a particle with mass
$m=\tilde{C}_b(\Phi_0/2\pi)^2$ in a potential
$U(\theta_b)=-E_b(\cos\theta_b+I_b\theta_b/I_r)$, $I_r=2\pi
E_b/\Phi_0$. For the biased case $I_b<I_r$, there exists a series of minima of $%
U(\theta_b)$, where $ \partial U(\theta_b)/\partial
\theta_b=0,\,\partial^2 U(\theta_b)/\partial \theta_b^2>0. $ Near
these points, $U(\theta_b)$ approximates to a harmonic oscillator
potential with a characteristic frequency
\begin{equation*}
\omega_b=\sqrt{\frac{2\pi I_r}{\tilde{C}_b\Phi_0}}\,\left[1-\left(\frac{I_b}{%
I_r}\right)^2\right]^{\,1/4}, 
\end{equation*}
depending on the applied bias-current $I_b$. Correspondingly, the
Hamiltonian $\hat{H}_r$ reduces to
\begin{equation}
\hat{H}_b=\left(\hat{a}^\dagger\hat{a}+\frac{1}{2}\right)\hbar\omega_b,
\end{equation}
with
\begin{equation*}
\hat{a}=\frac{1}{\sqrt{2}}\left[\left(\frac{\Phi_0}{2\pi}\right)\sqrt{\frac{%
\tilde{C}_b\omega_b}{\hbar}}\,\hat{\theta}_b+ i\left(\frac{2\pi}{\Phi_0}%
\right)\frac{\hat{p}_{\,b}}{\sqrt{\hbar\omega_b\tilde{C}_b}}\right],
\end{equation*}
and
\begin{equation*}
\hat{a}^\dagger=\frac{1}{\sqrt{2}}\left[\left(\frac{\Phi_0}{2\pi}\right)%
\sqrt{\frac{\tilde{C}_b\omega_b}{\hbar}}\,\hat{\theta}_b- i\left(\frac{2\pi}{%
\Phi_0}\right)\frac{\hat{p}_{\,b}}{\sqrt{\hbar\omega_b\tilde{C}_b}}\right].
\end{equation*}
The approximate number of quantum metastable bound
states~\cite{Martinis87} of the quantum oscillator is
$N_s=2^{3/4}\sqrt{E_b/E_{C_b}}(1-I_b/I_r)^{5/4}$ .

The energy scale of the quantum oscillator (3) is $\omega_b/(2\pi)\sim 10\,%
\mathrm{GHz}$~\cite{Martinis02}, which is of the same order of the
Josephson energy in the SQUID. Therefore, the oscillating quantum
of the information bus will be really excited, even if only one of
the qubits is operated quantum mechanically. This is different
from the case considered in \cite{Makhlin99}, wherein the
$LC$-oscillator shared by all charge qubits are not really
excited, as the eigenfrequency of the $LC$-circuit is much higher
than the typical frequencies of the qubits dynamics. For
operational convenience, we assume that the bus is coupled to only
one qubit at a time. The coupling between any one of the qubits
(e.g., the $k$th one) and the bus can, in principle, be controlled
by adjusting the applied external flux (e.g., $\Phi_k$). In this
case, any direct interaction does not exist between the qubits,
and the dynamics of the CBJJ can be safely restricted to the
Hilbert space spanned by the two Fock states: $|0_b\rangle$ and
$|1_b\rangle$, which are the lowest two energy eigenstates of the
harmonic oscillator of Eq. (3). Furthermore, we assume that the
applied gate voltage of any chosen ($k$th) qubit works near its
degeneracy point with $n_{g_k}=1/2$, and thus only
two charge states: $|n_k=0\rangle=|\uparrow_k\rangle$ and $%
|n_k=1\rangle=|\downarrow_k\rangle$, play a role during the
quantum operation. All other charge states with a higher energies
can be safely ignored. Therefore, the Hamiltonian
\begin{equation}
\hat{H}_{kb}\,=\,\hat{H}_k+\hat{H}_b+\lambda_k\left(\hat{a}^\dagger+\hat{a}%
\right)\sigma^{(k)}_y,
\end{equation}
with
\begin{equation}
\hat{H}_k\,=\,\left[\frac{\delta E_{C_k}}{2}\,\sigma^{(k)}_z-\frac{E_{J_k}}{2}%
\,\sigma^{(k)}_x\right],
\end{equation}
describes the interaction between any one of the qubits (e.g., the
$k$th one) and the bus, and provides the basic dynamics for the
present network. Here, $\delta E_{C_k}=2e^2(1-2n_{g_k})/C_k,\,\,
\lambda_k=C_{g_k}(2\pi/\Phi_0)\sqrt{\hbar/(2\tilde{C}_b\omega_b)}/(2C_k)$,
and the pseudospin operators are defined by:
\begin{equation*}
\left\{
\begin{array}{l}
\sigma^{(k)}_x\,=\,|$$\uparrow_k$$\rangle\langle$$\downarrow_k$$|+|$$\downarrow_k$$\rangle%
\langle$$\uparrow_k$$|, \\
\sigma^{(k)}_y\,=\,-i|$$\uparrow_k$$\rangle\langle$$\downarrow_k$$|+i|$$\downarrow_k$$
\rangle\langle$$\uparrow_k$$|, \\
\sigma^{(k)}_z\,=\,|$$\uparrow_k$$\rangle\langle
$$\uparrow_k$$|-|$$\downarrow_k$$\rangle\langle$$\downarrow_k$$|.%
\end{array}
\right.
\end{equation*}
Above, when the first cosine-term in Hamiltonian (2) was expanded, only the
single-quantum transition process approximated to the first-order of $\hat{%
\theta}_b$ was considered. The higher order nonlinearities have
been neglected as their effects are very weak. In fact, for the
lower number
states of the bus, we have $C_{g_k}\sqrt{ \langle\,\theta^2_b\,\rangle}%
/C_k\,\lesssim 10^{-2}$, for the typical experimental
parameters~\cite{Makhlin01,Tsai03,Wallraff03,Martinis03}: $C_b\sim
1pF,\,\,\omega_b/2\pi\sim 10\,{\rm GHz}$, and $C_{g_k}/C_{J_k}\sim
10^{-2}$.

Notice that the coupling strength $\lambda_k$ between the qubit
and the bus is tunable by controlling the flux $\Phi_k$, applied
to the selected qubit, and the bias-current $I_b$, applied to the
information bus. For example, such a coupling can be simply turn
off by setting the flux $\Phi_k$ as $\Phi_0/2$. This allows
various elemental operations for quantum manipulations to be
realizable in a controllable way. In the logic basis
$\{|0_k\rangle,|1_k\rangle\}$, defined by
\begin{equation*}
|0_k\rangle=\frac{|\downarrow_k\rangle+|\uparrow_k\rangle}{\sqrt{2}},\,\,
|1_k\rangle=\frac{|\downarrow_k\rangle-|\uparrow_k\rangle}{\sqrt{2}},
\end{equation*}
and under the usual rotating-wave approximation, 
the above Hamiltonian (4) can be rewritten as 
\begin{eqnarray}
\hat{H}_{kb}&=&\left[\frac{E_{J_k}}{2}\tilde{\sigma}^{(k)}_z- \frac{\delta
E_{C_k}}{2}\tilde{\sigma}^{(k)}_x\right] +\hbar\omega_b\left(\hat{a}^\dagger%
\hat{a}+\frac{1}{2}\right)  \notag \\
&+&i\lambda_k\left[\hat{a}\tilde{\sigma}^{(k)}_+\,-\hat{a}^\dagger\tilde{%
\sigma}^{(k)}_- \right],
\end{eqnarray}
with
\begin{equation*}
\left\{
\begin{array}{l}
\tilde{\sigma}^{(k)}_x\,=\,|1_k\rangle\langle0_k|+|0_k\rangle\langle1_k|,\\
\tilde{\sigma}^{(k)}_y\,=\,-i|1_k\rangle\langle0_k|+i|0_k\rangle\langle1_k|,\\
\tilde{\sigma}^{(k)}_z\,=\,|1_k\rangle\langle1_k|-|0_k\rangle\langle0_k|,%
\end{array}
\right.
\end{equation*}
and $\tilde{\sigma}^{(k)}_\pm=(\tilde{\sigma}^{(k)}_x\pm\tilde{\sigma}%
^{(k)}_y)/2$. Here, the logic states $|0_k\rangle$ and $|1_k\rangle$
correspond to the clockwise and anticlockwise persistent circulating
currents in the $k$th SQUID-loop, respectively.

We now discuss the quantum dynamics of the above Josephson network. Without
loss of generality, we assume in what follows that the bias-current $%
I_b$ applied to the CBJJ doesn't change, once it is set up
properly beforehand. The quantum evolutions of the system are then
controlled by other external parameters: the fluxes applied to the
qubits and the voltages across the gate capacitances of the
qubits. Depending on the different settings of the controllable
external parameters, different Hamiltonians can be induced from
Eq. (6) and thus different time-evolutions are obtained.
Obviously, during any operational delay $\tau$ with
$\Phi_{X_i}=\Phi_0/2$ and $V_k=e/C_{g_i}$, the $i$th qubit remains
in its idle state because the Hamiltonian vanishes (i.e.,
$H_0^{(i)}=0$) as $E_{J_i}=0,\,n_{g_i}=0$. However, the data bus
still undergoes a free time-evolution:
\begin{equation}
\hat{U}_0(t)=\exp\left(\frac{-it}{\hbar}\hat{H}_b\right).
\end{equation}
This evolution is useful for controlling the dynamical phase of
the qubits to exactly realize certain quantum operations. For the
other cases, the dynamical evolutions of the chosen qubit depend
on the different settings of the experimental parameters.

1) For the case where $\Phi_k=\Phi_0/2$ and $V_k\neq e/C_{g_k}$, the $i$th
qubit and the bus separately evolve with the Hamiltonians $\hat{H}%
_1^{(k)}=-\delta E_{C_k}\tilde{\sigma}^{(k)}_x/2$ and $\hat{H}_b$
determined by Eq. (3), respectively. The relevant time-evolution
operator of the whole system reads
\begin{eqnarray}
\hat{U}_1^{(k)}(t)&=&\exp\left(\frac{-it}{\hbar}\hat{H}^{(k)}_1\right)%
\otimes \exp\left(\frac{-it}{\hbar}\hat{H}_b\right).
\end{eqnarray}

2) If the $k$th qubit works at its degenerate point and couples to
the bus, i.e., $V_k=e/C_{g_k}$ and $\Phi_k\neq \Phi_0/2$, then we
have the Hamiltonian
\begin{equation}
\tilde{\hat{H}}_{kb}=E_{J_i}\tilde{\sigma}^{(k)}_z/2 +\hat{H}_b +i\lambda_k%
\left[\hat{a}\tilde{\sigma}^{(k)}_+\,-\hat{a}^\dagger\tilde{\sigma}^{(k)}_- %
\right]
\end{equation}
from (6). The corresponding dynamical evolutions are
\begin{widetext}
\begin{eqnarray}
\left\{
\begin{array}{ll}
|0_b\rangle|0_k\rangle\xrightarrow{\tilde{\hat{U}}_{kb}}
e^{i\Delta_kt/2}|0_b\rangle|0_i\rangle,\,
\tilde{\hat{U}}_{kb}=\exp(-i\tilde{\hat{H}}_{kb}t),\,\Delta_k=E_{J_k}/\hbar-\omega_b,
&  \\
\\
|0_b\rangle|1_k\rangle\xrightarrow{\tilde{\hat{U}}_{kb}}
e^{-i\omega_bt}
\left\{\left[\cos\left(\frac{\Omega_k}{2}t\right)-i\frac{
\Delta_k}{\Omega_k}\sin\left(\frac{\Omega_k}{2}t\right)\right]
|0_b\rangle|1_k\rangle-\frac{2\lambda_k}{\hbar\Omega_k}\sin\left(\frac{
\Omega_k}{2}t\right)|1_b\rangle|0_k\rangle\right\}, &  \\
\\
|1_b\rangle|0_k\rangle\xrightarrow{\tilde{\hat{U}}_{kb}}
e^{-i\omega_bt}
\left\{\left[\cos\left(\frac{\Omega_k}{2}t\right)+i\frac{
\Delta_k}{\Omega_k}\sin\left(\frac{\Omega_k}{2}t\right)\right]
|1_b\rangle|0_k\rangle+\frac{2\lambda_k}{\hbar\Omega_k}\sin\left(\frac{
\Omega_k}{2}t\right)|0_b\rangle|1_k\rangle\right\}, &
\end{array}
\right.
\end{eqnarray}
\end{widetext}
with $\Omega_k=\sqrt{\Delta^2_k+(2\lambda_k/\hbar)^2}$.

Specifically, we have the time-evolution operator
\begin{widetext}
\begin{eqnarray}
\hat{U}^{(k)}_{2}(t)=\hat{A}(t)\left(
\begin{array}{cc}
\cos\left(\frac{2\lambda_kt}{\hbar}\sqrt{\hat{n}+1}\right)
&-\frac{1}{\sqrt{\hat{n}+1}}
\sin\left(\frac{2\lambda_kt}{\hbar}\sqrt{\hat{n}+1}\right)\hat{a}\\
\\
\frac{\hat{a}^\dagger}{\sqrt{\hat{n}}}
\sin\left(\frac{2\lambda_kt}{\hbar}\sqrt{\hat{n}}\right)&
\cos\left(\frac{2\lambda_kt}{\hbar}\sqrt{\hat{n}}\right)\\
\end{array}
\right),
\end{eqnarray}
\end{widetext}
with $$\hat{A}(t)=\exp\left[-it\left(\frac{\hat{H}_b}{\hbar}
+\frac{E_{J_k}\tilde{\sigma}_z^{(k)}}{2\hbar}\right)\right],$$ for
the resonant case: $\Delta_k=0$. This reduces Eq. (10) to the time
evolutions:
\begin{widetext}
$$
\left\{
\begin{array}{ll}
|0_b\rangle|0_k\rangle\,\xrightarrow{\hat{U}^{(k)}_2(t)}\,|0_b\rangle|0_k\rangle,
&  \\
\\
|0_b\rangle|1_k\rangle\,\xrightarrow{\hat{U}^{(k)}_2(t)}e^{-i\omega_bt}\,
\left[ \cos\left(\frac{\lambda_kt}{\hbar}\right)
|0_b\rangle|1_k\rangle-\sin\left(
\frac{\lambda_kt}{\hbar}\right)|1_b\rangle|0_k\rangle\right],&  \\
\\
|1_b\rangle|0_k\rangle$$\,\xrightarrow{\hat{U}^{(k)}_2(t)}e^{-i\omega_bt}\,
\left[ \cos\left(\frac{\lambda_kt}{\hbar}\right)
|1_b\rangle|0_k\rangle+\sin\left(
\frac{\lambda_kt}{\hbar}\right)|0_b\rangle|1_k\rangle\right]. &
\end{array}
\right.
$$
\end{widetext}

For another extreme case, i.e., the system works in the dispersive regime
(far from the resonant point): $2\lambda_k/(\hbar|\Delta_k|)\ll 1 $, we have
the time evolution operator
\begin{eqnarray}
\tilde{\hat{U}}^{(k)}_{2}(t)=\hat{A}(t)\exp\left(-i\frac{\tilde{\hat{H}}%
^{\prime}_{kb}t}{\hbar}\right),
\end{eqnarray}
with
$$
\tilde{\hat{H}} ^{\prime}_{kb}=\lambda^2_k(|1_k\rangle\langle 1_k|\hat{%
a}\hat{a}^\dagger-|0_k\rangle\langle 0_k|\hat{a}^\dagger\hat{a}%
)/(\hbar\Delta_k).
$$
It reduces to the following time evolutions:
\begin{widetext}
$$
 \left\{
\begin{array}{ll}
|0_b\rangle|0_k\rangle\,\xrightarrow{\tilde{\hat{U}}^{(k)}_2(t)}\,\exp\left(it
\frac{\Delta_i}{2}\right)|0_b\rangle|0_i\rangle, &  \\
\\
|0_b\rangle|1_k\rangle\,\xrightarrow{\tilde{\hat{U}}^{(k)}_2(t)}\,
\exp\left[
-it\left(\omega_b+\frac{\Delta_i}{2}+\frac{\lambda^2_k}{\hbar^2\Delta_k}
\right)\right]|0_b\rangle|1_k\rangle, &  \\
\\
|1_b\rangle|0_k\rangle\,\xrightarrow{\tilde{\hat{U}}^{(k)}_2(t)}\,
\exp\left[
-it\left(\omega_b-\frac{\Delta_k}{2}-\frac{\lambda^2_k}{\hbar^2\Delta_k}
\right)\right]|1_b\rangle|0_k\rangle, &  \\
\\
|1_b\rangle|1_k\rangle\,\xrightarrow{\tilde{\hat{U}}^{(k)}_2(t)}\,
\exp\left[
-it\left(2\omega_b+\frac{\Delta_k}{2}+\frac{2\lambda^2_k}{\hbar^2\Delta_k}
\right)\right]|1_b\rangle|1_k\rangle. &
\end{array}
\right.
$$
\end{widetext}

3) Generally, if $\Phi _{k}\neq \Phi _{0}/2$ and $V_{g_{k}}\neq
e/C_{g_{k}}$, then the Hamiltonian (6) can be rewritten as
\begin{equation}
\bar{\hat{H}}_{kb}=\frac{E_{k}}{2}\bar{\sigma}_{z}^{(k)}+\hat{H}%
_{b}+i\lambda _{k}(\hat{a}^{\dagger }\bar{\sigma}_{-}^{(k)}-\hat{a}\bar{%
\sigma}_{+}^{(k)}),
\end{equation}%
with
$$
\left\{
\begin{array}{l}
 \bar{\sigma}_{x}^{(k)}=-\sin \eta _{k}\,\tilde{\sigma}_{z}^{(k)}-\cos\eta
_{k}\,\tilde{\sigma}_{x}^{(k)},\\
\bar{\sigma}_{y}^{(k)}=-\tilde{\sigma}_{y}^{(k)},\\
\bar{\sigma}_{z}^{(k)}=\cos\eta
_{i}\,\tilde{\sigma}_{z}^{(k)}-\sin\eta _{k}\,
\tilde{\sigma}_{x}^{(k)},
\end{array}
\right.
$$
and $\bar{\sigma}_{\pm }^{(k)}=(\bar{\sigma}_{x}^{(k)}\pm \bar{\sigma}%
_{y}^{(k)})/2$. Here, $\cos \eta _{k}=E_{J_{k}}/E_{k}$, and $E_{k}=\sqrt{%
(\delta E_{C_{k}})^{2}+E_{J_{k}}^{2}}$. If the bias-current
$I_{b}$ and the flux $\Phi _{k}$ are set properly beforehand such
that $E_{J_{k}}\sim
\hbar \omega _{b}\ll \delta E_{C_{k}}$, then the detuning $\hbar\bar{%
\Delta}_{k}=E_{k}-\hbar\omega _{b}$ is very large (compared to the coupling
strength $\lambda _{k}\lesssim 10^{-1}E_{J_{k}}$). Therefore, the
time-evolution operator of the system can be approximated as
\begin{eqnarray}
\bar{\hat{U}}^{(k)}_{3}(t)=\hat{B}(t)
\exp\left\{-i\frac{\lambda^2_kt}{\hbar^2\bar{\Delta}_k}
\left[\bar{\sigma}^{(k)}_z\left(\hat{a}^\dagger\hat{a}+\frac{1}{2}\right)+\frac{1}{2}\right]\right\},
\end{eqnarray}
with
$$
\hat{B}(t)=\exp\left[-it\left(\frac{\hat{H}_b}{\hbar}
+\frac{E_k\bar{\sigma}^{(k)}_z}{2\hbar}\right)\right].
$$
This implies the following evolutions
\begin{widetext}
$$
\left\{
\begin{array}{ll}
|0_{b}\rangle |0_{k}\rangle \xrightarrow{\bar{\hat{U}}^{(k)}_3(t)}%
e^{-i\zeta_kt}\left\{ [\cos (\xi_{k}t)+i\cos \eta _{k}\sin
(\xi_{k}t)]\,|0_{b}\rangle |0_{k}\rangle +i\sin \eta _{k}\sin
(\xi_{k}t)\,|0_{b}\rangle |1_{k}\rangle \right\} , &  \\
\\
|0_{b}\rangle |1_{k}\rangle \xrightarrow{\bar{\hat{U}}^{(k)}_3(t)}%
e^{-i\zeta_kt}\left\{ [\cos (\xi_kt)-i\cos \eta _{k}\sin
(\xi_kt)]\,|0_{b}\rangle |1_{k}\rangle +i\sin \eta _{k}\sin
(\xi_kt)\,|0_{b}\rangle |0_{k}\rangle \right\} , &  \\
\\
|1_{b}\rangle |0_{k}\rangle \xrightarrow{\bar{\hat{U}}^{(k)}_3(t)}%
e^{-i(\zeta_k+\omega _{b})t}\left\{ \left[ \cos (\xi_k^{\prime
}t)+i\cos \eta _{k}\sin (\xi_k^{\prime }t)\right] \,|1_{b}\rangle
|0_{k}\rangle +i\sin \eta
_{k}\sin (\xi_k^{\prime }t)\,|1_{b}\rangle |1_{k}\rangle \right\} , &  \\
\\
|1_{b}\rangle |1_{k}\rangle \xrightarrow{\bar{\hat{U}}^{(k)}_3(t)}%
e^{-i(\zeta_k+\omega _{b})t}\left\{ i\sin \eta _{k}\sin
(\xi_k^{\prime }t)\,|1_{b}\rangle |0_{k}\rangle +[\cos
(\xi_k^{\prime }t)-i\cos \eta _{i}\sin (\xi_k^{\prime
}t)]\,|1_{b}\rangle |1_{k}\rangle \right\}, &
\end{array}
\right.
$$
\end{widetext}
with
$$\zeta_k=\omega _{b}/2+\lambda _{k}^{2}/(2\hbar
^{2}\bar{\Delta}_{k})\,,\xi_{k}=E_{k}/(2\hbar )+\lambda
_{k}^{2}/(2\hbar ^{2}\bar{\Delta} _{k}),$$ and $$\xi_{k}^{\prime
}=\xi_{k}+\lambda _{k}^{2}/(\hbar ^{2} \bar{\Delta}_{i}).$$

In what follows we shall show that any process for manipulating the quantum
information stored in the present circuit can be effectively implemented by
selectively using the above elementary time-evolutions: $\hat{U}_0(t),\,\hat{%
U}_1^{(k)}(t),\,\hat{U}_2^{(k)}(t)$, $\tilde{\hat{U}}_2^{(k)}(t)$, and $\bar{%
\hat{U}}_3^{(k)}(t)$.

\section{Quantum manipulations of the superconducting nanocircuit.}

It is well known that any valid quantum transformation can be
decomposed into a sequence of elementary one- and two-qubit
quantum gates. The set of these gates is universal, and any
quantum computing circuit comprises only gates from this set.
Several schemes~\cite{Makhlin99,You02,Strauch03} have been
proposed for implementing one of the universal two-qubit gates
with Josephson qubits by using the direct interactions between
them. By making use of the data bus interacting sequentially with
the selective qubits, Blais \textit{et al.}~\cite{Blais03} showed
that the two-qubit gate may be effectively realized. Two important
problems will be solved in our indirect-coupling approach:

i) when one of two qubits is selected to couple with the data bus,
how we can let the remainder qubit decouple completely from the
bus; and

ii) the phase changes of the bus' and qubit's states during the
operations are very complicated, how we can control these phase
changes in order to precisely implement the desired quantum gate.

The scheme in~\cite{Blais03} assumed that, when one of the two
qubits is tuned to resonance with the bus, then the other qubit is
hardly affected because of its different Rabi frequency.
Obviously, this decoupling is not complete and thus it is not easy
to assure that the bus couples only one qubit at a time. By
controlling the external flux $\Phi_k$ applied to the qubits, the
network proposed here provides an effective method for making the
remainder qubit completely decouple from the bus. All the desired
elementary operations for quantum computing can be exactly
implemented by properly setting the experimentally controllable
parameters, e.g., the external $\Phi_k$, the gate voltage $V_k$,
the bias-current $I_b$, and the duration $t$ of each selected
quantum evolution, etc.

Hereafter, we assume that each of the selected time-evolutions can
be switched on/off very quickly.

\subsection{single-qubit operations}

First, we show how to realize the single-qubit operations on each
SQUID-qubit. This will be achieved by simply turning on/off the relevant
experimentally controllable parameters. For example, if $n_{g_k}\neq 1/2$
and $E_{J_k}=0$ for a time span $t$, then the time-evolution $\hat{U}%
^{(k)}_1(t)$ in equation (8) is realized. This operation is the
single-qubit rotation around the $x$ axis:
\begin{equation}
\hat{R}^{(k)}_x(\varphi_k)=\left(
\begin{array}{cc}
\cos\frac{\varphi_k}{2} & i\sin\frac{\varphi_k}{2} \\
\\
i\sin\frac{\varphi_k}{2} & \cos\frac{\varphi_k}{2}
\end{array}
\right),
\end{equation}
with $\varphi_k=\delta E_{C_k}t/\hbar$. Rotations by
$\varphi_i=\pi$ and $\varphi_k=\pi/2$ produce a spin flip (i.e., a
NOT-gate operation) and an equal-weight superposition of logic
states, respectively.

The rotation around the $z$ axis can be implemented by using the
evolution (12). This operation is conditional and dependent on the
state of the bus. If the bus is in the ground state $|0_b\rangle$,
the rotation reads
\begin{eqnarray}
R^{(k)}_z(\phi_k)=e^{-i\varrho_k t} \left(
\begin{array}{cc}
e^{-i\phi_k} & 0 \\
0 & e^{i\phi_k}%
\end{array}
\right),
\end{eqnarray}
with $\varrho_k=\omega_b/2+\lambda_k^2/(2\hbar^2\Delta_k),\,
\phi_k=E_{J_k}t/(2\hbar) +\lambda_k^2t/(2\hbar^2\Delta_k)$. With a
sequence of $x$- and $z$-rotations, any rotation on the
single-qubit can be performed. For example, the Hadamard gate
applied to the $k$th qubit:
\begin{equation*}
\hat{H}_{g}=\frac{1}{\sqrt{2}}\left(
\begin{array}{ll}
1 &\,\,\,\, 1 \\
1 & -1
\end{array}
\right),
\end{equation*}
can be implemented by a three-step rotation:
\begin{equation}
\hat{R}^{(k)}_z\left(\frac{\pi}{4}\right)\otimes\hat{R}^{(k)}_x\left(-\frac{\pi}{2}%
\right)\otimes\hat{R}^{(k)}_z\left(\frac{\pi}{4}\right) \,=\,-ie^{-i\kappa_i(t_1+t_3)}%
\hat{H}^{(k)}_{g}.
\end{equation}
Here, the relevant durations $t_1,\,t_2$, and $t_3$ are set
properly to satisfy the conditions
\begin{eqnarray*}
\cos\left(\frac{\delta E_{C_k}t_2}{\hbar}\right)&=&-\sin\left(\frac{\delta
E_{C_k}t_2}{\hbar}\right)  \notag \\
&=&\sin\left[\frac{E_{J_k}t_1}{2\hbar}+\frac{ (\lambda_k/\hbar)^2t_1}{%
2\Delta_k}\right]  \notag \\
&=&\sin\left[\frac{E_{J_k}t_3}{2\hbar }+\frac{(\lambda_k/\hbar)^2t_3}{%
2\Delta_k}\right]  \notag \\
&=&\frac{1}{\sqrt{2}}.
\end{eqnarray*}

\subsection{two-qubit operations}

Second, we show how to realize two-qubit gates by letting a pair
of qubits (the $k$th- and $j$th ones) interact separately with the
bus. Before the quantum operation, the chosen qubits decouple from
the bus. At the end of the desired gate operation the bus should
be disentangled again from the qubits, and returned to its ground
state. For operational simplicity, we assume that the bus
resonates with the control qubit, the $k$th one, i.e.,
$\Delta_k=0$. We now consider the following three-step operational
process:

i) Couple the control qubit to the bus (i.e., the applied external flux $%
\Phi_k$ is varied to $\Phi_0$) and realize the evolution $\hat{U}%
^{(k)}_1(t_1)$ for the duration $t_1$:
\begin{equation}
\sin\left(\frac{\lambda_k\,t_1}{\hbar}\right)=-1.
\end{equation}
Then, by returning the $\Phi_k$ to its initial value, i.e., $\Phi_k=\Phi_0/2$%
, the $k$th qubit can be decoupled from the bus exactly. Before
the next step operation, there is an operational delay $\tau_1$.
During this delay the state of the qubits does not evolve, while
the data bus still undergoes a time-evolution $\hat{U}_0(\tau_1)$.

ii) Couple the target qubit (the $j$th one) to the bus and realize
the time-evolution $\bar{\hat{U}}^{(j)}_3(t_2)$. This is achieved
by letting the chosen qubit work near its degenerate point (i.e.,
$n_{g_j}\neq 1/2$) and switching on its Josephson energy (i.e.,
$\Phi_j\neq \Phi_0/2$). After the time $t_2$ determined by the
condition
\begin{equation}
\cos(\xi_jt_2)=-\sin(\xi^{\prime}_jt_2)=1,
\end{equation}
we decouple the $j$th qubit from the bus and let it be in the idle
state by returning its gate-voltage $V_j$ to the degenerate point
($n_{g_j}=1/2$), and simultaneously switching off the relevant
Josephson energy. During another operational delay $\tau_2$ before
the next step operation, the bus undergoes another free-evolution
$\hat{U}_0(\tau_2)$.

iii) Repeat the first step and realize the evolution
$\hat{U}^{(k)}_1(t_3)$ with
\begin{equation}
\sin\left(\frac{\lambda_k\,t_3}{\hbar}\right)=1.
\end{equation}
Diagrammatically, the above three-step operational process with two delays
can be represented as follows:
\begin{widetext}
$$
\left\{
\begin{array}{lll}
|0_b0_k0_j\rangle\xrightarrow{\hat{U}_0(\tau_1)\hat{U}^{(k)}_2(t_1)}
e^{-i\omega_b\tau_1/2}|0_b0_k0_j\rangle
\xrightarrow{\hat{U}_0(\tau_2)\bar{\hat{U}}^{(j)}_3(t_2)}
e^{-i\chi}|0_b0_k0_j\rangle
\xrightarrow{\hat{U}^{(k)}_2(t_3)}e^{-i\chi}|0_b0_k0_j\rangle,\\
\\
|0_b0_k1_j\rangle\xrightarrow{\hat{U}_0(\tau_1)\hat{U}^{(k)}_2(t_1)}e^{-i\omega_b\tau_1/2}|0_b0_k1_j\rangle
\xrightarrow{\hat{U}_0(\tau_2)\bar{\hat{U}}^{(j)}_3(t_2)}e^{-i\chi}|0_b0_k1_j\rangle
\xrightarrow{\hat{U}^{(k)}_2(t_3)}e^{-i\chi}|0_b0_k1_j\rangle,\\
\\
|0_b1_k0_j\rangle\xrightarrow{\hat{U}_0(\tau_1)\hat{U}^{(k)}_2(t_1)}e^{-i\omega_b(t_1+3\tau_1/2)}|1_b0_k0_j\rangle\\
\hspace{3cm}\xrightarrow{\hat{U}_0(\tau_2)\bar{\hat{U}}^{(j)}_3(t_2)}ie^{-i\chi-i\omega_b(t_1+t_2+\tau_1+\tau_2)}
(\cos\eta_j|1_b0_k0_j\rangle+\sin\eta_j|1_b0_k1_j\rangle)\\
\hspace{5cm}\xrightarrow{\hat{U}^{(k)}_2(t_3)}ie^{-i\chi-i\omega_bT}
(\cos\eta_j|0_b1_k0_j\rangle+\sin\eta_j|0_b1_k1_j\rangle),\\
\\
|0_b1_k1_j\rangle\xrightarrow{\hat{U}_0(\tau_1)\hat{U}^{(k)}_2(t_1)}e^{-i\omega_b(t_1+3\tau_1/2)}|1_b0_k1_j\rangle\\
\hspace{3cm}\xrightarrow{\hat{U}_0(\tau_2)\bar{\hat{U}}^{(j)}_3(t_2)}ie^{-i\chi-i\omega_b(t_1+t_2+\tau_1+\tau_2)}
(\sin\eta_j|1_b0_k0_j\rangle-\cos\eta_j|1_b0_k1_j\rangle)\\
\hspace{5cm}\xrightarrow{\hat{U}^{(k)}_2(t_3)}ie^{-i\chi-i\omega_bT}
(\sin\eta_j|0_b1_k0_j\rangle-\cos\eta_j|0_b1_k1_j\rangle),\\
\end{array}
\right.
$$
\end{widetext}
with 
$T=t_{1}+t_{2}+t_{3}+\tau _{1}+\tau _{2}$ 
being the total duration of the process, and $\chi
=\zeta_{j}t_{2}+\omega _{b}(\tau _{1}+\tau _{2})/2$. Obviously,
the information bus remains in its ground state $|0_{b}\rangle $
after the operations. If the total duration $T$ is satisfied as
\begin{equation}
\sin (\omega _{b}T)=1,
\end{equation}
the above three-step process with two delays yields a two-qubit
gate
expressed by the following matrix form 
\begin{eqnarray}
\hat{U}^{(kj)}_1(\eta_j)=\left(
\begin{array}{cccc}
1 & 0 & 0 & 0 \\
0 & 1 & 0 & 0 \\
0 & 0 & \cos\eta_j & \sin\eta_j \\
0 & 0 & \sin\eta_j & -\cos\eta_j
\end{array}
\right),
\end{eqnarray}
which is a universal two-qubit Deutsch gate~\cite{Barenco95}.

Analogously, if the second step operation $\bar{\hat{U}}_{3}^{(j)}(t_{2})$
in the above three-step process is replaced by the operation $\tilde{\hat{U}}%
_{2}^{(j)}(t_{2})$, then another two-qubit operation expressed by
\begin{eqnarray}
\bar{\hat{U}}^{(kj)}_2(t_2)=\left(
\begin{array}{cccc}
\Gamma_j & 0 & 0 & 0 \\
0 & \Gamma^*_j & 0 & 0 \\
0 & 0 & \Lambda_j e^{-i\omega_bT} & 0 \\
0 & 0 & 0 & \Lambda^*_je^{-i\omega_bT}
\end{array}
\right),
\end{eqnarray}
with 
$\Gamma_j=\exp(i\varsigma_jt_2),\Lambda_j=\exp(i\varsigma'_jt_2),\,\varsigma_j=E_{J_{j}}/(2\hbar
)+\lambda _{j}^{2}/(2\hbar ^{2}\Delta
_{j}),\,\varsigma'_j=\varsigma_j+\lambda_{j}^{2}t_2/(\hbar ^{2}\Delta_{j})$, 
can be implemented. This three-step operational process can similarly be
represented diagrammatically as
\begin{widetext}
$$
\left\{
\begin{array}{lll}
|0_b0_k0_j\rangle\xrightarrow{\hat{U}_0(\tau_1)\hat{U}^{(k)}_2(t_1)}e^{-i\omega_b\tau_1}|0_b0_k0_j\rangle
\xrightarrow{\hat{U}_0(\tau_2)\tilde{\hat{U}}^{(j)}_2(t_2)} \Gamma
e^{-i\nu}|0_b0_k0_j\rangle
\xrightarrow{\hat{U}^{(k)}_2(t_3)}\Gamma e^{-i\nu}|0_b0_k0_j\rangle,\\
\\
|0_b0_k1_j\rangle\xrightarrow{\hat{U}_0(\tau_1)\hat{U}^{(k)}_2(t_1)}e^{-i\omega_b\tau_1}|0_b0_k1_j\rangle
\xrightarrow{\hat{U}_0(\tau_2)\tilde{\hat{U}}^{(j)}_2(t_2)}\Gamma^*
e^{-i\nu}|0_b0_k1_j\rangle
\xrightarrow{\hat{U}^{(k)}_2(t_3)}\Gamma^* e^{-i\nu}|0_b0_k1_j\rangle,\\
\\
|0_b1_k0_j\rangle\xrightarrow{\hat{U}_0(\tau_1)\hat{U}^{(k)}_2(t_1)}e^{-i\omega_b(t_1+3\tau_1/2)}|1_b0_k0_j\rangle
\xrightarrow{\hat{U}_0(\tau_2)\tilde{\hat{U}}^{(j)}_2(t_2)}\Lambda
e^{-i\nu-i\omega_b(t_1+t_2+\tau_1+\tau_2)}|1_b0_k0_j\rangle\\
\hspace{8.6cm}\xrightarrow{\hat{U}^{(k)}_2(t_3)}\Lambda e^{-i\nu-i\omega_bT}|0_b1_k0_j\rangle,\\
\\
|0_b1_k1_j\rangle\xrightarrow{\hat{U}_0(\tau_1)\hat{U}^{(k)}_2(t_1)}e^{-i\omega_b(t_1+3\tau_1/2)}|1_b0_k1_j\rangle
\xrightarrow{\hat{U}_0(\tau_2)\tilde{\hat{U}}^{(j)}_2(t_2)}\Lambda^*
e^{-i\nu-i\omega_b(t_1+t_2+\tau_1+\tau_2)}|1_b0_k1_j\rangle\\
\hspace{8.6cm}\xrightarrow{\hat{U}^{(k)}_2(t_3)}\Lambda^* e^{-i\nu-i\omega_bT}|0_b1_k1_j\rangle,\\
\end{array}
\right.
$$
\end{widetext}
with 
$\nu =\omega _{b}t_{2}/2+\lambda _{i}^{2}t_{2}/(2\hbar ^{2}\Delta
_{i})+\omega _{b}(\tau _{1}+\tau _{2})/2$. 
Above, the durations of the first- and third-step operations have been set
the same as those for realizing the two-qubit operation $\hat{U}%
_{1}^{kj}(\eta _{j})$.

The two-qubit gate $\hat{U}^{(kj)}_1(\eta_j)$ (or $\hat{U}^{(kj)}_2(t_2)$)
performed above forms a universal set. Any quantum manipulation can be
implemented by using one of them, accompanied by arbitrary rotations of
single qubits. Obviously, if the system works in the strong charge regime: $%
E_{J_{j}}/(\delta E_{C_{i}})\ll 1$, and $\cos \eta _{j}\sim 0,\,
\sin \eta _{j}\sim 1$, then the two-qubit gate
$\hat{U}^{(kj)}_1(\eta_j)$ in (22) approximates the well-known
controlled-NOT (CNOT) gate
\begin{equation*}
\hat{U}_{CNOT}^{(kj)}=\left(
\begin{array}{cccc}
1 & 0 & 0 & 0 \\
0 & 1 & 0 & 0 \\
0 & 0 & 0 & 1 \\
0 & 0 & 1 & 0%
\end{array}%
\right).
\end{equation*}
Also, if the duration $t_{2}$ of the evolution $ \tilde{\hat{U}}%
_{2}^{(j)}(t_{2})$ and the delays $\tau _{1},\,\tau _{2}$ are further set
properly such that
\begin{equation*}
\cos (\varsigma _{j}\,t_{2})=\sin (\varsigma _{j}^{\prime }t_{2})=\sin
(\omega _{b}T)=1,\,\
\end{equation*}
then the two-qubit operation $\hat{U}_{2}^{(kj)}$ in (23) reduces
to the well-known controlled-phase (CROT) gate
\begin{equation*}
\hat{U}_{CROT}^{(kj)}=\left(
\begin{array}{cccc}
1 & 0 & 0 & 0 \\
0 & 1 & 0 & 0 \\
0 & 0 & 1 & 0 \\
0 & 0 & 0 & -1%
\end{array}%
\right) .
\end{equation*}

\section{Decoherence of the qubit-bus system due to the biased voltage- and
current-noises}

An ideal quantum system preserves quantum coherence, i.e., its
time evolution is determined by deterministic reversible unitary
transformations. Quantum computation requires a long phase
coherent time-evolution. In practice, any physical quantum system
is subject to various disturbing factors which destroy phase
coherence. In fact, solid-state systems are very sensitive to
decoherence, as they contain a macroscopic number of degrees of
freedom and interact with the environment. However, coherent
quantum manipulations of the qubits are still possible if the
decoherence time is finite but not too short. Hence, it is
important to investigate the effects of the environmental noise on
the present quantum circuit.

The typical noise sources in Josephson circuits consist of the
linear fluctuations of the electromagnetic environments (e.g.,
circuitry and radiation noises) and the low-frequency noise due to
fluctuations in various charge/current channels (e.g., the
``background charge" and ``critical current" ). Usually, the
former one behaves as Ohmic dissipation~\cite{Weiss99} and the
latter one produces a $1/f$ spectrum~\cite{Paladino02}. Within the
present work, we will consider the case of Ohmic dissipation due
to linear fluctuations of the external circuit parameters: the
bias-current $I_{b}$ applied to the CBJJ and the gate voltages
applied to the qubits. The effect of gate-voltage noise on a
single charge qubit and that of bias-current noise on a single
CBJJ has been discussed in \cite{Makhlin01,Weiss99} and in
\cite{Martinis03}, respectively. We now study these noises
together (see figure 2),
\begin{figure}[tbp]
\vspace{1.5cm} \includegraphics[width=18cm]{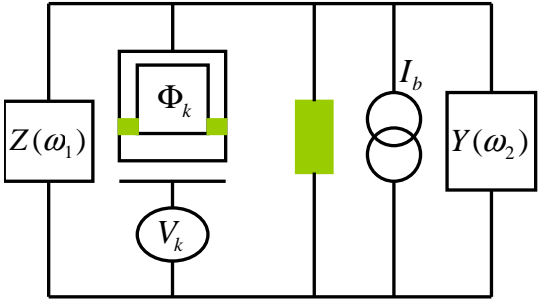}
\vspace{-9.2cm} \caption{Schematic diagram of a SQUID-based charge
qubit with impedance $Z(\protect\omega _{1})$ coupled to a CBJJ
with admittance $Y(\protect\omega _{2})$.}
\end{figure}
since the interaction between a CBJJ, acting as a bus here, and a
selected (e.g., the $k$th) qubit takes a central role in the
present scheme for quantum manipulations. Each electromagnetic
environment is treated as a quantum system with many degrees of
freedom and modeled by a bath of harmonic oscillators.
Furthermore, each of these oscillators is assumed to be weakly
coupled to the chosen system. The Hamiltonian of a chosen ($k$th)
qubit coupling to the bus, containing the fluctuations of the
applied gate voltage $V_{k}$ and bias-current $I_{b}$, can be
generally written as
\begin{equation*}
\hat{H}=\bar{\hat{H}}_{kb}+\hat{H}_{B}+\hat{V},
\end{equation*}%
with
\begin{eqnarray}
H_{B}&=&\sum_{j=1,2}\sum_{\omega _{j}}\left[ \frac{p_{\omega _{j}}^{2}}{%
2m_{\omega _{j}}}+\frac{m_{\omega _{j}}\omega _{j}^{2}x_{\omega _{j}}^{2}}{2}%
\right]  \notag \\
&=&\sum_{j=1,2}\sum_{\omega _{j}}\left( \hat{a}_{\omega _{j}}^{\dagger }\hat{%
a}_{\omega _{j}}+\frac{1}{2}\right) \hbar \omega _{j},
\end{eqnarray}
and
\begin{equation}
\hat{V}=-\left[\sin \alpha _{k}\overline{\sigma }_{z}^{(k)}+\cos \alpha _{k}%
\overline{\sigma }_{x}^{(k)}\right](\hat{R}_{1}+\hat{R}_{1}^{\dagger })-(%
\hat{a}^{\dagger }R_{2}+\hat{a}\hat{R}_{2}^{\dagger }),
\end{equation}%
being the Hamiltonians of the two baths and their interactions with the
non-dissipative qubit-bus system $\bar{\hat{H}}_{kb}$, respectively. Above, $%
\hat{a}_{\omega _{j}},\hat{a}_{\omega _{j}}^{\dagger }$ are the Boson
operators of the $j$th bath, and
\begin{equation*}
\hat{R}_{1}=\frac{eC_{g_{k}}}{C_{k}}\sum_{\omega _{1}}g_{\omega _{1}}\hat{a}%
_{\omega _{1}},\,\,\,\,\,R_{2}=\sqrt{\frac{\hbar }{2\widetilde{C}_{b}\omega _{b}}}%
\sum_{\omega _{2}}g_{\omega _{2}}\hat{a}_{\omega _{2}},
\end{equation*}%
with $g_{\omega _{j}}$ being the coupling strength between the
oscillator of frequency $\omega _{j}$ and the non-dissipative
system. The effects of these noises can be characterized by their
power spectra, which in turn depend on the corresponding
``impedance" (or ``inductance") and the temperature of the
relevant circuits. For example, introducing the impedance
$Z_{t}(\omega )=1/[i\omega C_{t}+Z^{-1}(\omega )]$ with $Z(\omega
)=R_{V}$ being the Ohmic resistor, the corresponding voltage
between the terminals of impedance $Z_{t}(\omega )$ can be
expressed as $\delta V=\sum_{\omega _{1}}\lambda _{\omega
_{1}}x_{\omega _{1}}$. Thus, the spectral density of this voltage
source for Ohmic dissipation can be expressed as
\begin{eqnarray}
G(\omega )&=&\pi \sum_{\omega _{1}}\frac{\lambda _{\omega
_{1}}^{2}}{2m\omega _{1}}\delta (\omega
-\omega _{1})\notag \\
&=&\pi \sum_{\omega_1}|g_{\omega _{1}}|^2\delta (\omega -\omega
_{1})\,\sim\,\,R_{V}\omega.
\end{eqnarray}%
Similarly, the spectral density for the bias-current source can be
approximated as
\begin{eqnarray}
F(\omega )=\pi \sum_{\omega_2}|g_{\omega _{2}}|^2\delta (\omega
-\omega _{2})\,\sim\,\, Y_{I}\omega,
\end{eqnarray}%
with $Y_{I}$ being the dissipative part of the admittance of the current
bias.

The well-established Bolch-Redfield
formalism~\cite{AK64,Goorden03} offers a systematic way to obtain
a generalized master equation for the reduced density matrix of
the system, weakly influenced by dissipative environments. A
subtle Markov approximation is also made in this theory such that
the resulting master equation is local in time. Of course, in the
regime of weak bath coupling and low temperatures, this theory is
numerically equivalent to a full non-Markovian path-integral
approach~\cite{Hartmann00}. For the present qubit-bus system and
in the basis spanned by the eigenstates $\{|g\rangle,|u_{n}\rangle
,|v_{n}\rangle ,n=1,2,...\}$ of the non-dissipative Hamiltonian
$\bar{\hat{H}}_{kb}$, the Bloch-Redfield theory leads to the
following master equations
\begin{equation}
\frac{d\sigma _{\alpha \beta }}{dt}=-i\omega _{_{\alpha \beta }}\sigma
_{_{\alpha \beta }}+\sum_{\mu ,\nu }\left( R_{\alpha \beta \mu \nu
}+S_{\alpha \beta \mu \nu }\right) \sigma _{\mu \nu },
\end{equation}%
with
\begin{widetext}
\begin{eqnarray}
R_{\alpha \beta \mu \nu } &=&-\frac{1}{\hbar ^{2}}\int_{0}^{\infty
}d\tau \times \left[g_{1}(\tau )\left( \delta _{\beta \nu
}\sum_{\kappa }A_{\alpha \kappa }A_{\kappa \mu }e^{i\omega _{\mu
\kappa }\tau }-A_{\alpha
\mu }A_{\nu \beta }e^{i\omega _{\mu \alpha }\tau }\right) \right. \nonumber\\
&+&\left. g_{1}(-\tau )\left( \delta _{\alpha \mu }\sum_{\kappa
}A_{\nu \kappa }A_{\kappa \beta }e^{i\omega _{\kappa \nu }\tau
}-A_{\alpha \mu }A_{\nu \beta }e^{i\omega _{\beta \nu }\tau
}\right) \right] ,
\end{eqnarray}
\end{widetext}
and
\begin{widetext}
\begin{eqnarray}
S_{\alpha \beta \mu \nu } &=&-\frac{1}{\hbar ^{2}}\int_{0}^{\infty
}d\tau \times \left[ g_{2}^{\dagger }(\tau )\left(\delta _{\beta
\nu }\sum_{\kappa }B_{\alpha \kappa }^{\dagger }B_{\kappa \mu
}e^{i\omega _{\mu \kappa }\tau }-B_{\alpha \mu }B_{\nu \beta
}^{\dagger }e^{i\omega _{\mu \alpha }\tau }\right)\right.\nonumber\\
&+&g_{2}^{\dagger }(-\tau )\left( \delta _{\alpha \mu
}\sum_{\kappa }B_{\nu \kappa }^{\dagger }B_{\kappa \beta
}e^{i\omega _{\kappa \nu }\tau
}-B_{\alpha \mu }B_{\nu \beta }^{\dagger }e^{i\omega _{\beta \nu }\tau }%
\right)\nonumber\\
&+&g_{2}^{-}(\tau )\left(\delta _{\beta \nu }\sum_{\kappa
}B_{\alpha \kappa }B_{\kappa \mu }^{\dagger }e^{i\omega _{\mu
\kappa }\tau }-B_{\alpha \mu }^{\dagger }B_{\nu \beta }e^{i\omega
_{\mu \alpha }\tau }\right)\nonumber\\
&+&\left.g_{2}^{-}(-\tau )\left( \delta _{\alpha \mu }\sum_{\kappa
}B_{\nu \kappa }B_{\kappa \beta }^{\dagger }e^{i\omega _{\beta
\kappa }\tau
}-B_{\alpha \mu }^{\dagger }B_{\nu \beta }e^{i\omega _{\beta \nu }\tau }%
\right)\right] ,
\end{eqnarray}
\end{widetext}
with
\begin{widetext}
\begin{eqnarray*}
\left\{
\begin{array}{l}
g_{1}(\pm \tau
)=\left(\frac{eC_{g_{k}}}{C_{k}}\right)^{2}\sum_{\omega
_{1}}|g_{\omega _{1}}|^{2}[\langle n(\omega _{1})+1\rangle e^{\mp
\,i\omega _{1}\tau }+\langle n(\omega _{1})\rangle e^{\pm
\,i\omega _{1}\tau }],\\
\\
g_{2}^{\dagger }(\pm \tau )=\left( \frac{\hbar
}{2\widetilde{C}_{b}\omega _{b}}\right) \sum_{\omega
_{2}}|g_{\omega _{2}}|^{2}\langle n(\omega _{2})+1\rangle e^{\mp
\,i\omega _{2}\tau },\\
\\
g_{2}^{-}(\pm \tau )=\left( \frac{\hbar }{2\widetilde{C}_{b}\omega
_{b}}\right) \sum_{\omega _{2}}|g_{\omega _{2}}|^{2}\langle
n(\omega _{2})\rangle e^{\mp \,i\omega _{2}\tau }.
\end{array}
\right.
\end{eqnarray*}
\end{widetext}%
Above, each one of the states $|\alpha\rangle,|\beta\rangle,...$
can be equal to one of the eigenstates of $\bar{\hat{H}}_{kb}$.
$\langle n(\omega _{j})\rangle =1/[\exp (\hbar \omega
_{j}/k_{B}T)-1]$ is the average number of thermal photons in the
mode of frequency $\omega _{j}$. The denotation $x_{ab}=\langle
\alpha|\hat{x}|\beta\rangle $ accounts for the matrix element of
operator $\hat{x}$, i.e.,
$$
A_{\alpha\beta}=\langle \alpha|\hat{A}_{k}|\beta\rangle ,\,\,\,
\hat{A}_{k}\,=\,\overline{\sigma }_{z}^{(k)}\sin \alpha
_{k}+\overline{\sigma }_{x}^{(k)}\cos \alpha _{k}\,=\,\sigma
_{z}^{(k)},
$$
and
$$B_{\alpha\beta}\,=\,\langle \alpha|\hat{a}|\beta\rangle
,\,\,\,B_{\alpha\beta}^{\dagger }\,=\,\langle
\alpha|\hat{a}^{\dagger }|\beta\rangle .
$$
Also, $\omega _{\alpha \beta }=(E_{\alpha }-E_{\beta })/\hbar $ with $%
E_{\alpha }\,\,(E_{\beta })$ being one of eigenvalues of the
non-dissipative Hamiltonian $\bar{\hat{H}}_{kb}$, corresponding to
the eigenstate $|\alpha \rangle $\thinspace (\thinspace $|\beta
\rangle $\thinspace ). The spectrum of $\bar{\hat{H}}_{kb}$
includes the ground state $|g\rangle
=|-_{k},0\rangle ,\,$corresponding to the energy $E_{g}=-\hbar \overline{%
\Delta }_{k}/2,$ and a series of dressed doubled states
\begin{equation*}
\left\{ \begin{array}{l} |u_{n}\rangle\, =\,\cos \theta
_{n}|+_{k},n\rangle -i\sin \theta _{n}|-_{k},n+1\rangle
,\\
\\
|v_{n}\rangle \,=\,-i\sin \theta _{n}|+_{k},n\rangle +\cos \theta
_{n}|-_{k},n+1\rangle
\end{array}
\right.
\end{equation*}%
corresponding to the eigenvalues
\begin{equation*}
E_{u_{n}}\,=\,\hbar \omega _{b}(n+1)-\frac{\rho
_{n}}{2},\,\,\,\,E_{v_{n}}=\hbar \omega _{b}(n+1)+\frac{\rho
_{n}}{2},
\end{equation*}%
with
$$\cos \theta _{n}\,=\,\rho _{n}-\hbar \overline{\Delta
}_{k}/\sqrt{(\rho _{n}-\hbar \overline{\Delta }_{k})^{2}+4\lambda
_{k}^{2}(n+1)},$$ and
$$
\rho _{n}\,=\,\sqrt{(\hbar
\overline{\Delta }_{k})^{2}+4\lambda _{k}^{2}(n+1)}.
$$
Here, $|\pm _{k}\rangle $ and $|n\rangle $ are the eigenstates of the operators $%
\overline{\sigma }_{z}^{(k)}$ and $\hat{H}_{b}$ with eigenvalues
$\pm 1$ and $\hbar \omega _{b}(n+1/2)$, respectively.

Under the secular approximation, the evolution of the non-diagonal element $%
\sigma _{\alpha \beta }$ of the reduced density matrix $\sigma $ is
determined by
\begin{eqnarray}
\frac{d}{dt}\sigma _{\alpha \beta }&+&\left\{ i\left[\omega _{\alpha \beta }+%
{\rm Im}(R_{\alpha \beta \alpha \beta })+{\rm Im}(S_{\alpha \beta
\alpha \beta })\right]\right.\nonumber\\
&+&\left.\left[{\rm Re}(R_{\alpha \beta \alpha \beta })+{\rm Re}%
(S_{\alpha \beta \alpha \beta })\right]\right\} \sigma _{\alpha
\beta }=0.
\end{eqnarray}%
Here, $R_{\alpha \beta \mu \nu }$ and $S_{\alpha \beta \mu \nu }$ are
calculated respectively from $R_{\alpha \beta \mu \nu }$ and $S_{\alpha
\beta \mu \nu }$ by setting $\mu =\alpha $ and $\nu =\beta $. ${\rm Re}%
(x) $ and ${\rm Im}(x)$ represent the real- and imaginary parts of
the complex number $x$. The formal solution of the above
differential equation (31) reads
\begin{eqnarray}
\sigma _{\alpha \beta }(t)\,=\,\sigma_{\alpha \beta
}(0)\exp\left(-T^{-1}_{\alpha \beta }t\right) \exp
\left(-i\Theta_{\alpha \beta}t\right),
\end{eqnarray}
with $\Theta_{\alpha \beta}=\omega_{\alpha \beta }+{\rm
Im}(R_{\alpha \beta \alpha \beta })+{\rm Im} (S_{\alpha \beta
\alpha \beta }) $ being the effective oscillating frequency (the
original Bohr frequency $\omega_{\alpha \beta }$ plus the Lamb
shift $\Delta\omega_{\alpha\beta}={\rm Im}R_{\alpha \beta \alpha
\beta }+{\rm Im} S_{\alpha \beta \alpha \beta }$), and
\begin{equation}
T_{\alpha \beta }^{-1}=-[{\rm Re}(R_{\alpha \beta \alpha \beta })+%
{\rm Re}(S_{\alpha \beta \alpha \beta })]
\end{equation}
describing the rate of decoherence between the states $|\alpha
\rangle $ and $|\beta \rangle $.

In the present qubit-bus system operating near the resonant point:
$E_{k}\sim\hbar \omega _{b}$, the decoherences relating to the
lowest three energy eigenstates, i.e., $|g\rangle $,\thinspace
$|u_{0}\rangle =|u\rangle $,\thinspace and $|v_{0}\rangle
=|v\rangle $, are specially important for the desired quantum
manipulations. The decoherences outside these three states are
negligible. After a long but direct derivation, we obtain the
decoherence rates of interest:
\begin{widetext}
\begin{eqnarray}
T_{gu}^{-1}&=&\alpha _{V}\left\{4\left(\sin \alpha _{k}\cos
^{2}\theta _{0}\right)^{2}\frac{2k_{B}T}{\hbar }+2\left(\cos
\alpha _{k}\cos \theta _{0}\right)^{2} \coth\left(\frac{\hbar
\omega _{ug}}{2k_{B}T}\right)\omega
_{ug}\right.\nonumber\\
&+&\left.\left(\cos\alpha _{k}\sin \theta
_{0}\right)^{2}\left[\coth\left(\frac{\hbar \omega _{vg}}{2k_{B}T}\right)-1\right]\omega _{vg}
+\left(\sin \alpha _{k}\sin 2\theta _{0}\right)^{2}\left[\coth \left(%
\frac{\hbar \omega _{vu}}{2k_{B}T}\right)-1\right]\omega
_{vu}\right\}\nonumber\\
 &+&\alpha _{I}\sin ^{2}\theta
_{0}\left\{\coth \left(\frac{\hbar \omega
_{ug}}{2k_{B}T}\right)+1\right\}\omega _{ug},
\end{eqnarray}
\begin{eqnarray}
T_{gv}^{-1} &=&\alpha _{V}\left\{4\left(\sin \alpha \sin
^{2}\theta _{0}\right)^{2}\frac{2k_{B}T}{\hbar }+2\left(\cos
\alpha \sin \theta _{0}\right)^{2}\coth \left(\frac{\hbar
\omega _{vg}}{2k_{B}T}\right)\omega _{vg}\right.\nonumber\\
&+&\left.\left(\cos \alpha \cos \theta _{0}\right)^{2}\left[\coth \left(\frac{\hbar \omega _{ug}}{2k_{B}T%
}\right)-1\right]\omega _{ug}
+\left(\sin \alpha \sin 2\theta _{0}\right)^{2}\left[\coth
\left(\frac{\hbar \omega _{vu}}{2k_{B}T}\right)+1\right]\omega
_{vu}\right\}\nonumber\\
&+&\alpha
_{I}\cos ^{2}\theta _{0}\left\{\coth \left(%
\frac{\hbar \omega _{vg}}{2k_{B}T}\right)+1\right\}\omega _{vg},
\end{eqnarray}%
\end{widetext}
and
\vspace{0.02cm}
\begin{widetext}
\begin{eqnarray}
T_{uv}^{-1} &=&\alpha _{V}\left\{4\left(\sin \alpha \cos 2\theta _{0}\right)^{2}\frac{%
2k_{B}T}{\hbar }+2\left(\sin \alpha \sin 2\theta
_{0}\right)^{2}\coth
\left(\frac{\hbar \omega _{vu}}{2k_{B}T}\right)\omega _{vu}\right.\nonumber\\
&+&\left.
\left(\cos \alpha \cos \theta _{0}\right)^{2}\left[\coth \left(\frac{\hbar \omega _{ug}}{2k_{B}T%
}\right)+1\right]\omega _{ug}+\left(\cos \alpha \sin \theta
_{0}\right)^{2}\left[\coth \left(\frac{\hbar \omega
_{vg}}{2k_{B}T}\right)+1\right]\omega
_{vg}\right\}\nonumber\\
&+&\alpha _{I}\left\{\sin ^{2}\theta _{0}\left[\coth \left(\frac{\hbar \omega _{ug}}{2k_{B}T}%
\right)+1\right]\omega _{ug}+\cos ^{2}\theta _{0}\left[\coth \left(\frac{\hbar \omega _{vg}}{%
2k_{B}T}\right)+1\right]\omega _{vg}\right\}.
\end{eqnarray}%
\end{widetext}
Above, the various Bohr frequencies read
$$\omega _{ug} =\omega _{b}/2+E_{k}/(2\hbar)-\sqrt{(\hbar \omega
_{b}-E_{k})^{2}+4\lambda _{k}^{2}}/(2\hbar),$$
$$\omega _{vg}
=\omega _{b}/2+E_{k}/(2\hbar)+\sqrt{(\hbar \omega
_{b}-E_{k})^{2}+4\lambda _{k}^{2}}/(2\hbar),$$
and
$$\omega _{vu}
=\sqrt{(\hbar \omega _{b}-E_{k})^{2}+4\lambda _{k}^{2}}/\hbar.$$
Two dimensionless parameters 
$\alpha _{V} =\pi
R_{V}C_{g_{k}}^{2}/[R_{K}C_{k}^{2}],\,R_{K}=h/e^{2}\approx
25.8~k\Omega$\, and $\alpha _{I} =Y_{I}/(\widetilde{C}_{b}\omega
_{b})$
characterize the coupling strengths between the environments and
the system.

Specially, if the system works far from the resonant point (with
$\lambda _{k}\sim 0,$ achieved by switching off the Josephson
energy), the above results (shown in Eqs. (34-36)) reduce to
those~\cite{Makhlin01,Martinis03,Weiss99} for the case when the
qubit and the bus independently decohere. Namely, $T_{gu}^{-1}$
reduces to the rate~\cite{Makhlin01}
$$
T_{\uparrow \downarrow }^{-1}=8\alpha _{V}k_{B}T/\hbar,
$$
which describes
the decoherence between two charge states $|\downarrow \rangle $ and $%
|\uparrow \rangle $ of the superconducting box with zero Josephson
energy. Also, $T_{gv}^{-1}$ reduces to the decoherent
rate~\cite{Martinis03}
$$
T_{01}^{-1}=\alpha _{I}[\coth (\hbar \omega _{b}/2k_{B}T)+1]\omega
_{b},
$$
between the ground and first excited states of the data bus.
However, for the strongest coupling case (i.e., when the system
works at the resonant point), we have $E_{k}=E_{J_{k}}=\hbar
\omega _{b}$,\,$\cos \alpha _{k}=1,\cos \theta _{0}=\sin \theta
_{0}=1/\sqrt{2}$,\, and
$\coth[\hbar \omega _{ug}/(2k_{B}T)]-1\simeq \coth [\hbar\omega
_{vg}/(2k_{B}T)]-1\,\sim\, 0$\,\,(\,$<10^{-7}$,\,
for the typical experimental parameters~\cite{Tsai99}: $\lambda
_{k}\simeq 0.1E_{J_{k}},\,E_{J_{k}}=\hbar \omega _{b}\simeq 50\mu
eV\gg k_{B}T\simeq 3\mu eV$). Thus, the minimum decoherent rates
\begin{eqnarray}
\widetilde{T}_{gu}^{-1} &=&(\alpha _{V}+\alpha _{I})\omega _{ug}, \\
\nonumber\\
\widetilde{T}_{gv}^{-1} &=&(\alpha _{V}+\alpha_{I})\omega _{vg},
\end{eqnarray}%
and
\begin{equation}
\widetilde{T}_{uv}^{-1}=\widetilde{T}_{gu}^{-1}+\widetilde{T}_{gv}^{-1},
\end{equation}
 are obtained for the above three dressed states, respectively.

It has been estimated in Ref.~\cite{Makhlin01} that the
dissipation for a single SQUID-qubit is sufficiently weak: $\alpha
_{V}\sim 10^{-6}$ for $R_V=50\Omega,\,C_{J_k}/C_{g_k}\sim
10^{-2}$, which allows, in principle, for $10^{6}$ coherent
single-qubit manipulations. For a single CBJJ the dimensionless parameter $%
\alpha _{I}$ only reaches $10^{-3}$ for typical experimental
parameters~\cite{Martinis02}: $1/Y_{I}\sim 100~\Omega $,
$C_{b}\sim ~6~p$F, $\omega _{b}/2\pi \,\sim 10$\,GHz. This implies
that the quantum coherence of the present qubit-bus system is
mainly limited by the bias-current fluctuations.
Fortunately, the impedance of the above CBJJ can be engineered~\cite{Martinis02} to be $%
1/Y_{I}\sim 560~$k$\Omega $. This lets $\alpha _{I}$ reach up to
$10^{-5}$ and allow about $10^{5}$ coherent manipulations of the
qubit-bus system.

\section{Conclusions and discussions}

In summary, we have proposed an effective scheme to couple any
pair of selective Josephson charge qubits by letting them
sequentially couple to a common CBJJ, which can be treated as an
oscillator with adjustable frequency. Two logic states of the
present qubit are encoded by the clockwise and anti-clockwise
persistent circuiting currents in the dc SQUID-loop. At most one
qubit can be set to interact with the bus at any moment. The
interaction between the selected qubit and the data bus is tunable
by controlling the flux applied to the qubit and the bias-current
applied to the data bus. This selective coupling provides a simple
way to manipulate the quantum information stored in the connected
SQUID-qubits. Indeed, any pair of selective qubits without any
direct interaction can be entangled by using a three-step coupling
process. Furthermore, if the total duration is set up properly,
the desired two-qubit universal gates, which are very similar to
the CNOT- and CROT gates, can be implemented via such three-step
operational processes. During this operation, the mode of the data
bus is unchanged, although its vibrational quantum is really
excited/absorbed. After the desired quantum operation is performed
on the chosen qubits, the data bus disentangles from the qubits
and returns to its ground state.

In previous schemes, the distant Josephson qubits are coupled
directly by either the charge-charge interaction, via connecting
to a common capacitor, or by a current-current interaction, via
sharing a common inductor. The present indirect coupling scheme
offers some advantages: i) the coupling strength is tunable and
thus easy to be controlled for realizing the desired quantum gate,
ii) this first-order interaction is more insensitive to the
environment, and thus possesses a longer decoherence time. Also,
compared to previous data buses, the externally connected
$LC$-resonator~\cite{Falci03} and cavity QED mode~\cite{You03},
the present CBJJ bus might be easier to control for coupling the
chosen qubit. For example, its eigenfrequency can be controlled by
adjusting the applied dc bias-current. In addition, the CBJJ is
easy to fabricate using current technology~\cite{Ramos01} and may
provide more effective immunities to both charge and flux noise.

By considering the decoherence due to the linear fluctuations of
the applied voltage $V_k$ and current $I_b$, we have analyzed the
experimental possibility of the present scheme within the
Bloch-Redfield formalism. A simple numerical estimate showed that
the quantum manipulations of the present qubit-bus system are
experimentally possible, once the impedance $Y_I$ of the CBJJ can
be engineered to have a sufficient low value, i.e., $1/Y_{I}$ can
be enlarged sufficiently (e.g., $1/Y_{I}\sim 560$\,K $\Omega$
\cite{Martinis02}). Of course, this possibility, like those in
previous schemes~\cite{Makhlin99,Falci03,You02,You03,Blais03}, is
also limited by other technological difficulties, e.g., suppress
the low-frequency $1/f$ noise, and fast switch on/off the external
flux to couple/decouple the chosen qubit, etc.. For example, a
very high sweep rate of magnetic pulse (e.g., up to $\sim\,
10^{8}$ Oe/s~\cite{A}), is required to change half of flux quantum
through a SQUID-loop (with the size e.g., $50\mu{\rm m}$) in a
sufficiently short time (e.g., the desired $\sim 40$ ps). This and
other obstacles pose a challenge that motivate the exploration of
novel circuit designs that might minimize some of the problems
that lie ahead in the future.

\section*{Acknowledgments}
This work was supported in part by the National Security Agency
(NSA) and Advanced Research and Development Activity (ARDA) under
Air Force Office of Research (AFOSR) contract number
F49620-02-1-0334, and by the National Science Foundation grant No.
EIA-0130383.

\end{document}